\newif\ifdraft\drafttrue
\documentclass[screen, nonacm]{acmart}

\renewcommand\footnotetextcopyrightpermission[1]{} 
\setcopyright{none} 
\makeatletter
\let\@authorsaddresses\@empty
\makeatother

\usepackage[utf8]{inputenc}
\usepackage{fullpage}
\usepackage{natbib}
\usepackage{xcolor}
\definecolor{darkred}{HTML}{8b0000}
\definecolor{darkgreen}{HTML}{006400}
\definecolor{darkblue}{HTML}{00008b}
\definecolor{darkmagenta}{HTML}{3f003f}
\usepackage{hyperref}
\hypersetup{
    colorlinks=true,            
    linkcolor=darkred,          
    citecolor=darkgreen,        
    filecolor=darkblue,         
    urlcolor=darkmagenta        
}

\begin{document}

\title{Report on the ``The Future of the Shell'' Panel at HotOS 2021}

\author{Michael Greenberg}
\authornote{Denotes alphabetical ordering and equal contribution.}
\affiliation{
  \institution{Stevens Institute of Technology}
  \country{USA}
}
\email{mgreenbe@stevens.edu}

\author{Konstantinos Kallas}
\authornotemark[1]
\affiliation{
  \institution{University of Pennsylvania}
  \country{USA}
}
\email{kallas@seas.upenn.edu}

\author{Nikos Vasilakis}
\authornotemark[1]
\affiliation{
  \institution{Massachusetts Institute of Technology}
  \country{USA}
}
\email{nikos@vasilak.is}

\author{Stephen Kell}
\affiliation{
  \institution{King's College London}
  \country{UK}
}
\email{srkell@acm.org}

\begin{abstract}
  This document summarizes the challenges and possible research
  directions around the shell and its ecosystem, collected during and after the HotOS21
  Panel on the future of the shell. The goal is to create a snapshot
  of what a number of researchers from various disciplines---connected
  to the shell to varying degrees---think about its future.
  We hope that this document will serve as a reference for future
  research on the shell and its ecosystem.
\end{abstract}

\maketitle
\fancyhf{} 

\section{What was ``The Future of the Shell'' Panel?}

Michael Greenberg, Konstantinos Kallas, and Nikos Vasilakis organized and ran a panel titled “The Future of the Shell: Unix and Beyond” at HotOS XVIII on June 3rd, 2021, from 1pm to 2:30pm ET~\cite{shellpanel:21}. The entire event was virtual due to COVID-19, held on Zoom (with support from Slack and a shared Google Doc for notes).

The goal of the event was to suggest the shell as an area worthy of attention and research, using discussions to find themes and problems.
As a conversation starter, the event was closely coupled to a position paper published by the panel's organizers in the same venue and presented the day before, highlighting key recent enablers and open problems in the shell~\cite{jash:21}.
By bringing together researchers and engineers from disparate communities---e.g., systems, languages, security---the event aimed to explore and highlight impactful research directions around the shell. 

The event was only loosely a panel. After an introduction by the organizers, four panelists---Arjun Guha from Northeastern University, Deepti Raghavan from Stanford University, Chet Ramey from Case Western Reserve University, and Diomidis Spinellis from the Athens University of Economics and Business and the Delft University of Technology---spoke briefly on their various interests in the shell. After these statements, over 100 attendees dispersed into three breakout rooms to discuss. Each breakout room had an organizer (to take notes and orchestrate the discussion) and one or two of the four panelists.

The breakout room sessions lasted roughly 45 minutes, followed by a five-minute break. After the break, all attendees gathered in a single virtual room to summarize discussions and share insights.

\section{What did people talk about?}

The discussion was loose and far-ranging in each of the three rooms. We can group discussion into a few broad categories:
usability and other human factors (Section~\ref{sec:usability}), expressiveness (Section~\ref{sec:expressiveness}), performance (Section~\ref{sec:performance}), and legacy and compatibility (Section~\ref{sec:legacy}).
The report is entirely in the past tense. The narrative presented here is our summary of the conversation; the structure we've introduced here is from our post hoc analysis. While some opinions are attributed, many assertions in the text here were made by one or more attendees, with the possibility of additional mistakes in understanding or transcription on our part.
Our hope is that this document stimulates further research---it may occasionally do so by being wrong!

We've organized our discussion following those four themes, with each such section ending with an ``Outlook'' subsection with discussion (or simply questions) for possible future work.
But first we summarize our panelists' opening remarks.

\subsection{Opening remarks}
\label{sec:opening}

In their opening remarks, the four panelists drew attention to the convenience and power of the shell... while also making arguments about why the shell is often not the right tool for the job---for example, if the goal is performance,  modern data formats, safety, or usability. Diomidis highlighted that the shell is the common language of computing systems, as a command-line interface, scripting language, and as a program's API.
While the shell remains a \emph{lingua franca}, the shell is also a language-agnostic tool designed to work in harmony with other languages---including domain-specific languages such as \texttt{sed} and \texttt{awk} but also arbitrary executables. Unfortunately, string manipulation is the primary glue; finding better approaches for language-agnostic composition is an important research direction.

Arjun identified the crux of the shell's issues as follows: the shell isn't always the right tool for the job, and the web is awash with guides on rewriting quick-and-dirty shell scripts into more robust, full-fledged languages (e.g., Python). There are many benefits in starting with a shell script---use of existing tools, interactive development, developer economy---but at some point developers will need to switch/port/rewrite their program to a `saner' programming language. How could we help make this rewriting easier?

Diomidis pointed out that the shell's technical power is somewhat lost, as the shell is failing as a socio-technical system. He identified three core issues:
the shell lacks systematic instruction---educationally the shell is an afterthought, rarely formally taught;
POSIX conventions, manpages, and the like are on the retreat;
and the shell and its ecosystem has stagnated, with poor support for modern data formats, programming in the large, and distribution.
The shell, Diomidis says, has missed out on improvements in performance, efficiency, expressiveness, consistency, and reliability.

Deepti highlighted prior work showing that shell scripts can enjoy these modern improvements: with appropriate metadata for individual commands, we can apply analyses and transformations.
She cited parallelization (PaSh~\citep{pash}) and distribution (POSH~\citep{posh}) in particular, but there are many interesting research directions along this line. With proper support, the shell could make use of the heterogeneous cloud resources common in modern deployments.
Relatedly, most distributed systems are designed to tolerate failures---but with the shell language and commands it is near impossible to be able to tolerate (distributed) faults, as there is no clear insight into the state and effects of programs involved in shell scripts and thus there is no clear understanding of what needs to be recovered in order to re-compute a result correctly. Therefore an important research direction is to design distributed fault-tolerance mechanisms for the shell and make recovering computations significantly easier.

Chet Ramey focused on the engineering perspective. In \texttt{bash}, he has experimented with several features, some of which were better received than others.
He's maintained a high level of POSIX compatibility along with significant extensions in what is the most usable shell for a wide variety of users, expectations, and expertise---ranging from a highly customizable interactive environment to a broad set of features to a programming language that is accessible by many users. Backwards compatibility, he emphasized, is a real concern.

\section{Usability and other human factors}
\label{sec:usability}

Shell scripts are hard even for experts---there's an awful lot of Googling and StackOverflow involved! Why is that? Can we improve it?
Discussants brought up several key issues:
command-line shells typically have no affordances (in the sense of Gibson~\cite{gibson2014ecological});
the terminal-oriented user interface misses out on modern improvements and can be hostile to newcomers;
shell programming is rarely formally taught;
and shell use varies significantly between user classes and application domains.
In addition to these themes, several rooms brought up the possibility of program synthesis for the shell.

\subsection{Affordances}

When one opens a shell, nothing happens at all: just a blank prompt.
Irene Zhang, Elena Glassman, and others noted that the shell offers no \emph{affordances} (in the sense of Gibson~\cite{gibson2014ecological})---if you don't know how to operate something just by the looks of it, you're doomed.
The shell is often used for quick and dirty data processing and exploration for text files. Experienced shell users know several commands and their flags and are quickly able to patch together scripts that achieve their goals, but less experienced users are forced to search for solutions on Google or StackOverflow, or read through long \texttt{man} pages that contain a lot of information that they don't need.
Offering more affordances would certainly improve shell pedagogy (see below).

Elena Glassman pointed out that the shell one-liner replacing over 100 lines of Java mentioned in the accompanying HotOS21 paper by the panel's organizers~\cite{jash:21} is, in fact, unreadable by non-experts. The script uses commands that are between two and four characters long---and if you know them that's great, but if you don't know them then there's no easy way to reconstruct them. Such complex commands are a significant learning cliff, and the Java program might in fact be more understandable. 
She related the concept to so-called Norman doors~\cite{norman1988psychology}, whose affordances do not match the direction of the door (e.g., a pull door with a pushbar rather than a handle). Autocompletion is a possible (partial) solution to such misleading affordances: names can be long and descriptive, but can be automatically completed by the system (commonly using \verb|<TAB>|).
Shorter names can be fine aliases for experts who want to type it quickly, but, to build affordances into the shell, the default could be using longer descriptive names. Experts and non-experts alike could enjoy the benefits of speed, but non-experts would also enjoy the benefits of affordances.\footnote{Textual interfaces aren't the only ones with such issues---voice interfaces like Alexa or Siri don't offer a menu of things one can ask.}
While PowerShell is often derided as being verbose, it is more readable and offers a partial solution to affordances with its verb-noun metaphor e.g., \texttt{remove drive}, which is reminiscent of the arguments in favor scripting languages like, say,  AppleScript.
Chet Ramey points out that this might be more of a naming convention issue rather than a shell usability issue, at least in the case of reading one's own code. Shell users can leverage aliases to make their own code more readable. Nikos Vasilakis has noted a challenge similar to the one expressed by Elena when teaching shell programming to undergraduates, who have indicated that names like \texttt{grep} and \texttt{sed} are too cryptic---and has considered an idea similar to the one from Chet: develop a small library of aliases such as \texttt{filter} (for \texttt{grep}) and \texttt{line-transform} (for \texttt{sed}) that would be \texttt{source}d at the beginning of any shell script. A key challenge here is that each command provides a DSL tailored for a specific set of use cases---so giving specialized commands generic names may in fact make it more confusing for newcomers: what are the differences between \texttt{sed}-like \texttt{transform} and \texttt{tr}-like \texttt{transform} and \texttt{awk}-like \texttt{transform}?

\subsection{Terminals and GUIs}

Some attendees were more interested in the shell-as-a-user-interface---how people organize and invoke things---and less interested in the shell-as-a-programming-language. David Karger was interested in identifying which aspects of this dual nature are fundamental to the shell and which are accidents. For constrast, a lot of the tasks one performs on the shell---such as moving files around---can be easily achieved using GUIs nowadays (e.g., files-and-folder interfaces) and have the affordances that are missing from the shell. The cost for the user is that they are then restricted to operate within the very limited vocabulary that these interfaces have. For example, the Finder and Spotlight applications on macOS often obscure the hierarchical structure of folders, leaving novices confused about where their files actually are.
Andy Chu's Oil shell~\citep{oil} attempts, in part, to decouple the shell from the terminal, communicating with the terminal via a socket. The GUI is excellent at showing state, and thus the GUI can query the shell and show the state as affordances or hints. You can still type to the shell, but one can always display state above or below the user's input line, or on the side.
David Holland suggested a windowing environment for the Unix shell---i.e., a windowing environment tailored to the shell (rather than just a multiplexer); he cited TempleOS and its hyperlinks as relevant.
A GUI restricts the use of language with direct affordances versus a blank terminal. What would Scratch for the shell look like? How would it represent the available moves (i.e., what are the affordances)?

Another approach would be to overlay affordances on top of the shell by developing run-time reflection and introspection capabilities à la Smalltalk. Many commands already support a form of introspection thanks to manual pages or help text, both available in quasi-standard ways (via the \texttt{man} command or using a \texttt{-h} option). However, these fall short of a metaprogramming system because they provide only human-readable text, not a structured self-description that could be the basis of further tool support.
Manual pages also frequently feature examples, which can be more illuminating than general descriptions. This suggests an obvious complement to completion systems: whereas completion is about filling in details, examples are about templates capturing the primary usages, in the expectation that users will fill in their own details.
Even so, a variety of tools parse \texttt{man} pages to automatically generate shell completions.

\subsection{Pedagogy}

How do people learn to use the CLI? Why do some people stick with it while others avoid it? What about elitism?
University students nowadays do not often reach for the shell for their everyday tasks but rather go straight to Python and various notebooks. Their use of the shell tends to be one-off and StackOverflow-driven.

The novices' experience is significantly different from the experience of senior researchers, who often turn to the shell for everyday tasks. What are the reasons that have led to this shift? More importantly, which of these reasons are accidental, and which are fundamental? Python and Jupyter notebooks are the best tool for several tasks, e.g., ML and statistics experiments, but do not scale and cannot be composed effectively with a bigger system. For these (and other reasons), experts often resort to shell scripts. Why don't students reach for the shell in problems where it is the right solution? What can we learn from notebooks?

Three main causes were identified during the discussion. First, Python is taught much more widely. It is perhaps the most popular introductory programming language.
Students naturally reach for the most familiar tool.
Second, notebooks provide a significantly easier-to-use interface than shell scripts (especially for beginners). Younger students do not feel comfortable with text-based interfaces (since most commercial technology has shifted away from them) and therefore need to overcome that challenge too before using the shell. Notebooks offer an integrated environment with a descriptive visual interface that can be easily navigated and used appropriately. Python also has rich debugging and profiling support that also improves user experience for everyone (novices and professionals alike).
Finally, the shell depends on (and therefore requires understanding of) Unix and the filesystem abstraction. To use the shell, one needs to already know about directories, the current directory, files, file descriptors, links, etc. Again, commercial technology is continuously shifting away from these abstractions, hiding the file system from plain sight, and providing different interfaces for navigation, search, and management of data. For example, mobile phone operating systems (and some personal-computer operating systems) try really hard to hide the filesystem and file abstraction, instead offering data-management capabilities through specific applications (a photo gallery for images and other visual content, a music application for music and other audio content, etc).
Users are accustomed to many affordances and very little concrete information about how their system is organized; the shell is the opposite, offering too much information at once, but without affordances entirely.

Diomidis Spinellis also highlighted the problem as a failure of the shell as sociotechnical system, with bad (or nonexistent) pedagogy being a culprit.
The shell is not systematically taught: most people learn it on the road, through a few incantations to get their job done, and from that point it's just searching. Proven knowledge and conventions are on the retreat, even among experts---for example Python's \texttt{argparse} library doesn't follow the POSIX standard and a command's documentation through the Unix manual is on the decline.

\subsection{User profiles and domains}

Timothy Roscoe and others pointed out that shell users can be separated into different categories, with different needs:
system administrators, application developers, DevOps and SRE engineers, experimenters, data scientists, and learners all need different things from their shell.
Depending on the user, different levels of system detail and complexity are appropriate.
Related, the differing domains call for different tools and approaches.
What problem is the shell trying to solve, today? And who is solving that problem? The shell is applied for conventional systems and software tasks, but also by scientists munging data or simply installing software, or by end-users copy-pasting in a command to, e.g., turn off Spotlight's file indexer on macOS.

As someone said at the panel, ``Back in the day everyone was a genius''. For example, \texttt{ed} gives ``\texttt{?}'' as its only error message, and people lived with it.
Bafflingly, TOPS20 was a contemporaneous and popular system, with a great interface and smart contextual help! Even experts appreciate nice user interfaces.

Shell scripts are often created as small, useful pieces of code to be shared with other people.
Reproducing a computational pipeline, and explaining it to somebody else, ought to be as easy as cloning or visualizing a Git repository.
Yet anecdotally shell scripts are often ``write once read never'', and considered hard to understand or maintain (per Elena Glassman, above).
Although it is possible to write bad code in any language, these tendencies may owe to technical and cultural factors worth investigating further.

As for domains of application: in which domains is the shell better than the alternatives? One such domain is describing workflows by composing applications, i.e., composing ``libraries'' from other languages. The shell is particular good at this because everything on the \texttt{PATH} is imported by default, and because files and FIFOs are convenient abstractions for communicating between these applications. That is, the shell is excellent glue.
But where else does the shell excel? Many data processing tasks that would have been the shell's purview thirty years ago are now done in Python or JavaScript. Libraries like Xonsh or Plumbum help narrow this gap further~\citep{xonsh,plumbum}.
Will future shell languages continue to be custom-built DSLs, or will we see systems that use general-purpose programming languages as their shell language?

\subsection{Synthesis}

Deepti Raghavan suggested example-based synthesis of shell commands: allow the developers to specify the input-output transformation and then have a system generate the set of shell commands required for achieving this transformation. Often shell programs are significantly shorter than their Python counterparts for achieving certain string transformations---but it's unclear which exact combination of commands and flags achieves the result one wants. An automated system---possibly around program synthesis---should be able to achieve this goal.
Giving examples for filesystem manipulations is more complex---perhaps something akin to Word macros and their `record' feature would be appropriate.
Tortoise is in this space: it can synthesize changes to a Vagrant script in response to user interactions~\cite{tortoise}.
Tom Van Cutsem and others noted that there is prior work and significant interest in the automated translation of natural language to shell commands. Most recently, IBM organized the NLC-to-CMD workshop\footnote{\url{http://nlc2cmd.us-east.mybluemix.net/}} which showed that achieving the aforementioned usability goals might be closer than we originally thought: while Deep Learning is very costly, NLP interfaces to the shell look like they will soon be available and tackling the aforementioned discoverability and affordances challenges. In a recent example, researchers asked a system trained on NLP-to-shell commands to ``destroy all PDF files" to which the system synthesized a pipeline piping the results of \texttt{find} to \texttt{shred} (rather than \texttt{rm}), illustrating that systems pick up on nuanced high-level semantics.
A similar problem has been identified in other domains, such as spreadsheets and database query engines, where synthesis using input-output examples has been a very promising solution (see Microsoft Flashfill). Could such techniques be applied in the domain of the shell, essentially automating shell invocations given a set of example input-output pairs?
Given the domain, synthesizing filesystem manipulations seems particularly dangerous---especially if the output of synthesis is hard to read and the user driving the synthesis is a novice! How can we make this synthesis safe, or its results explorable without commitment?

\subsection{Outlook}

How can we make the shell have (more) affordances or hints?
Is there low hanging fruit from our knowledge of good UI design that we can instantiate in existing shell environments?
How could one bridge the textual terminals and affordance-rich user interfaces?
How do you give people the power to invoke what's accessible by typing a few characters on the shell but with some of the ease and affordances in graphical user interfaces?
What is appropriate pedagogy for the shell? What can we do to support the shell as a robust sociotechnical system?
Understanding the domains in which the shell excels can help scope future work on the shell to focus on these domains, as well as allow us to port some of the shell's powerful characteristics to other languages and programming environments that can benefit from them.
Finally, synthesis is a promising direction for shell scripts---what would input/output examples look like? How would we ensure that the generated scripts were safe?

\section{Expressiveness}
\label{sec:expressiveness}

The shell is the main way that programmers and administrators control computers.
Some tasks are easier than others in the shell. Adding features to the shell and tools to its ecosystem can ease some of the pain of difficult tasks.
Attendees identified a few key challenges:
shortcomings for modern cloud deployments;
difficulty orchestrating complex patterns of communication;
messy cleanup and teardown of global state;
challenges handling modern formats;
and lack of support for migrating shell scripts to other languages.

\subsection{Distributed shell and cloud deployments}

Existing programming models are not adequate for working with distributed systems or in the cloud.
The shell is designed for a time-shared multi-user single computer setting; it is designed for executing interactive tasks more than batch processing.
Existing cloud offerings offer a suite of custom tools (some on the command line, some via a web interface) to manage resources.
What is an alternative/refined interface for interacting with a cluster of computers? Is there a uniform approach to setting up sessions? The underlying system would need to work transparently, even though nodes might crash or restart and the network might fail.
The problem of offering distribution-aware operating system abstractions for and at the level of the shell has remained open for at least forty years~\cite{catalogue} and is a recurring theme at HotOS~\cite{m31, jash:21}.

Arjun Guha pointed out that continuous integration (CI) frameworks like Travis allow one to develop programs where there is transparency over which computer they run on and whether it fails. But this transparency is achieved through volume/disk snapshots that have to be explicitly identified by the developer. This works but is very cumbersome---especially compared to having a simple pipe.

According to Timothy Roscoe, the shell is actually simple because it ignores all of the issues concerning distribution, failures, etc.
For example, picking the machine you want to run on is important for several use cases, e.g., at Google and other companies with heavy workloads---people who care to run each task on the right hardware for efficiency. Do we want to expose that complexity in a multi-machine shell, or do we want to hide it? The world is complex, so a shell should probably not hide that complexity but rather expose it.
Chet Ramey agreed: such simplicity ``is the shell sticking to what it does well''. We should put ``all of the mess of distributed computation and execution in separate utilities that the shell can compose''.

Doug McIlroy suggested that the shell should pursue `distributable computing'---a term coined by Victor Vyssotsky. Distributable computing is a paradigm of composing components together without worrying about where those components live. Such an approach as seen in Misra's Orc~\cite{orclang} makes sense for many cloud deployments, allowing communicating processes to run on separate hardware without their knowing it. Can a shell facilitate this, or must it run on a purpose-built operating system, like Plan 9?

\subsection{Complex patterns of communication}

The shell offers very simple communication and composition between processes (FIFOs, pipelines) as long as communication is one-to-one.
Chet Ramey pointed out that extensions (like process substitution) or new shells (like dgsh~\citep{dgsh}) make it easier.
The shell's computational model corresponds to processes (which are independent computation entities) and one-to-one channels for communication between them.
No solution offers clean ways to achieve many-to-one and one-to-many communication patterns; it is often much easier to resort to writing a program in a different language to achieve that. The \texttt{/dev/fd} functionality can help with scatter-gather type communication but it is not that convenient (and has other downsides; see Section~\ref{sec:formats}).

Doug McIlroy suggested developing novel combinators for complex communication patterns. Today, Bourne-descended shells smoothly provide only a few major combinators: ``\texttt{;}'' for sequential operation, ``\texttt{|}'' for functional composition, and ``\texttt{\&}'' for setting things off in the background.
Recent shells may support additional combinators---for example, dgsh supports combinators for expressing directed acyclic graphs (DAGs).
Unfortunately, many of these notations are clumsy, which raises two immediate questions: (1) is there good notation for useful combinators, and (2) which combinators are actually useful in practice?
Examples of useful combinators include the permutation of an array of streams and stream feedback, i.e., cyclic processes which feed outputs back as their own inputs.
The adoption of new combinators such as feedback should be motivated by examples of useful programs that exploit them. Nikos Vasilakis noted that feedback might be useful for a combined web-crawling-and-indexing pipeline and other worklist-type algorithms.

Tom Van Cutsem noted that colleagues at Bell Labs recently offered an elegant solution to introducing additional combinators---essentially fork/join-style parallelism for shell scripts~\cite{pbj}. Their system is named PB\&J from the names of the three key combinators Parallel, Branch, and Join, each of which is implemented with higher order commands named \texttt{p}, \texttt{b}, and \texttt{j}, respectively.

\subsection{Global state}

The shell's state is generally global: environment variables are dynamically scoped by default, and the filesystem is a global store shared by all running programs.
Modern general-purpose programming languages avoid such global state and shared memory since they often lead to races.
Is it possible to make shell commands more modular or, at least as a first step, infer bounds on how they affect and are affected by their environment? Could we infer specifications of the variables and files read or written by a command invocation?

Relatedly, it is hard to modularize shell scripts. Clean setup and teardown of runtime environments (temporary files, etc.) is a challenging problem, made more challenging by the shell's dynamism and the filesystem's shared nature.
A variety of systems address the problem at a variety of levels (e.g., \texttt{chroot}, \texttt{virtualenv}/\texttt{rvm}/etc., Docker, VMs), but none are perfectly adapted to the shell's needs here.

The shared global state's raciness can lead to serious nondeterminism---an unpleasant property for, e.g., build scripts! Some members of the Stanford SNR group asked what ecosystem support could  guarantee that each program's output is a deterministic function of its declared inputs. Given such deterministic tooling, the shell could help record ``why'' provenance for each output, borrowing ideas from the database literature.

\subsection{Modern formats}
\label{sec:formats}

The shell and its ecosystem excel at working with line-oriented formats, but it is less able to handle modern formats, like JSON or YAML or TOML or XML.
Discussants identified two approaches: ecosystem extensions and shell extensions.

In terms of extending the ecosystem, a variety of tools have emerged for various data formats: \texttt{jq} and \texttt{gron} work on JSON; \texttt{csvkit} (and especially its \texttt{csvsql} command) is good for CSV. \texttt{xmlstarlet} offers a similar suite of tools for XML.
Gabriel Weaver's \texttt{xutools} offer a more general approach, with a notion of grammar-specific path for working with structured formats.    

The shell itself can of course be extended.
Non-POSIX features such as arrays make it easier to process data with more structure. With \texttt{bash} as evidence, such structures can clearly be introduced without legacy breakage.
Shells also offer extension features, e.g., \texttt{bash} has a \texttt{csv} loadable builtin that lets shell scripts access CSV using array variables.
Shells could be extended with a much broader notion of data element (see, e.g., nushell, and Section~\ref{sec:legacy}). It is important that the ecosystem supports such notions, e.g., \texttt{head} should return the first $N$ data elements of a list, not just lines.
Clearly specifying the mappings between complex data structures and lists of strings will help adapt line-based tools somewhat, but not perfectly.

A more thoroughgoing approach might extend the ecosystem beyond mere tools, i.e., offering new operating system tools and facilities.
Userspace filesystems have a long history in Unix and Plan 9~\citep{Killian84procfs,Faulkner91procfs}.
Approaches like ffs~\citep{ffs} support multiple formats and let the
existing ecosystem work in its usual way on the directory mapping of a
file~\citep{Wimmer18fad}.
Tools like AVFS have taken this approach, and userspace filesystems are used to support a variety of tasks in GNOME.

Stephen Kell suggested that since the shell directly embeds objects from the underlying Unix---the environment and, especially, the filesystem---the conceptually cleanest place for extensions is often in the operating system beneath the shell, not within the shell itself.
He gave as an example that working with structured data within files, it would be possible to extend the shell, but better to expose that structure through the underlying filesystem.
The same general approach can apply to network services, code providers or `module'-like constructs, and so on, pursuing the classical Unix `everything is a file' design.
Chet Ramey noted that established system boundaries can make this difficult, not technically but politically---as a shell maintainer, adding features within the OS is out of reach.
As a workaround, filesystem-like features have been added in \texttt{bash}, such as the \texttt{/dev/tcp} and \texttt{/dev/fd} pseudo-files. These additions have been controversial, as they usurp OS prerogatives.
All told, these sorts of extension highlight the lack of coordination noted in Diomidis Spinellis's opening remarks.

\subsection{Migrating scripts}

While some shell scripts are long-lived, many programs start as shell scripts but move to more robust languages.
The reasons are varied: more efficient data structures and other performance concerns; tighter integration to existing systems, maintainability, and other software engineering concerns; and correctness and security concerns.
There is very little support for such migrations. What tools would help?

Can we use a core language (like, say CoLiS~\citep{colis}, or something in the spirit of $\lambda_{JS}$ \cite{lambdajs}) to describe shell scripts in a lower-level, shell-independent abstraction (e.g. invocation of a process with particular arguments, environment, file descriptors, and filesystem state)?
If so, we can probably use higher-level tools to reason about the execution of shell-invoked pipelines, and evolve the shell's language and UI independently of its behavior.
The Oil shell~\cite{oil} is an attempt in this direction, separating the user-facing language from the core execution engine.

\subsection{Outlook}

What do the ``separate utilities'' for managing ``all of the mess of distributed computation and execution'' look like, and how can the shell compose them?
How transparently can such distribution be achieved? What level of heterogeneity can we support transparently?
What are some other real-life examples of programs in the shell's domain that use feedback? What other communication primitives and combinators would be useful? What suite of tasks or corpus of shell scripts would make a good benchmark for communication patterns?
What tools can make clean setup and teardown easier for the shell? How can we help with introspection and debuggability? How can we help write deterministic shell scripts?
How can we extend the shell's expressiveness in worthwhile, impactful ways without getting in the operating system's way? Which extensions belong in the shell, and which belong outside of it? What are low-cost ways to extend the shell's ecosystem so that \emph{existing} tools continue to work?
What would a core language for shell look like?

\section{Performance}
\label{sec:performance}

In general, the time and space performance of shell scripts is determined by how the commands composing the shell script behave.
While some aspects of shells can be sped up---e.g., Andy Chu is working towards minimizing forking in the Oil shell~\citep{oil}---improving the performance of shell scripts will mean improving command compositions.
Savvy compositions of commands---manually or by an optimizer---can indeed lead to huge speedups through, e.g., data parallelism and other concurrency \citep{pash,dgsh} or distribution \citep{posh}.

Beyond questions of `raw' performance, there are other concerns, namely scalability and incrementality.
Ting Dai noted distributed modes for shells are key for scalability (e.g., \texttt{pssh}, POSH).
As for incrementality, shell scripts are often used for the preparation of an execution environment (downloading and installing dependencies) and for data (pre)processing tasks. These scripts have two common characteristics: (i) they take a long time to complete (often regardless of the computational resources since they are bound by things like network throughput), and (ii) they are often developed in an interactive fashion, since it is difficult to know the target end result \emph{a priori}, requiring small incremental changes. Unfortunately, after each of these changes, the user usually needs to rerun the whole script (or comment out some initial part of it to not rerun it). 
Incremental computation support would improve the process: if part of the script (or its input) changes, re-execution need not happen for the whole script and data, but only for the modified portions and their dependencies.
Incrementality already works in heavily constrained domains, e.g., Makefiles. But broader shell script support does not exist, and adding it will bring the usual legacy and compatibility issues (Section~\ref{sec:legacy}).
Chet Ramey classed incrementality as a modularity issue: ``build your pipeline as a set of independent scripts (or, more clumsily, shell functions) that can be tested individually''.
Doug McIlroy disagreed with the `clumsiness' of the function approach:
``it may also be clumsy to understand or maintain a bunch of very
short related scripts that are required to inhabit separate files''.
Software engineering advice is good, but tooling that helps you do the right thing is often better!

\subsection{Outlook}

Existing work has pushed towards improved performance through data parallelism (\citep{pash, odfm, dgsh}) and distribution (\citep{posh,gg}).
Approaches in this area rely on annotations of commonly used commands, which the optimizers/planners use to rearrange programs. Developing these annotations is arduous, and errors can corrupt programs more or less arbitrarily. Worse still, each project uses its own annotations! A common language here would be an enabling technology.
Incrementality is less well understood: what command annotations are necessary? What role does virtualization have to play?

\section{Legacy and compatibility}
\label{sec:legacy}

The shell is specified as part of POSIX~\cite{posix} (\cite{POSIXbase}, Section ``Shell \& Utilities'').
Different platforms---Linux, macOS, the BSDs, and Windows---treat the POSIX standard and specifications differently. Some of these differences are explicitly allowed for by marking certain behaviors implementation-defined, unspecified, or undefined; other differences are simply divergences from the standard.
As such, the shell is a commonly used tool where users might observe platform-specific behaviors.

The availability of \texttt{bash} on these platforms simplifies things somewhat; for example, Google mandates \texttt{bash} as the shell to use~\cite{googlebash}. As Chet Ramey observed, ``If you can count on having a particular shell everywhere you want to run your script (with particular version and feature requirements), you don't have to be compatible with anything''. Even so, demanding \texttt{bash} in all circumstances doesn't work for everyone.
Build scripts---whether for managing packages or testing software in continuous integration (CI) environments---are ideally portable. Accordingly, ``portable shell'' is an important but loosely understood concept.

What are the important considerations for moving the shell forward? The POSIX standard doesn't really innovate---they standardize what existing shells do.
Discussants identified two angles here: updating the POSIX spec and reimagining the shell.

On the one hand, the POSIX spec is far too loose. For example, local variables are left entirely unspecified, even though shells exhibit a very narrow range of behaviors. Legislating a single behavior will be hard, though: as Chet Ramey said, “it all comes down to whose ox is being gored”.
Various shell implementors and other interested parties continue to refine the POSIX spec to reflect changes in the shell landscape. For example, the draft for the next edition of the POSIX spec includes the \texttt{pipefail} shell option, which allows for better diagnosis of pipeline failures.\footnote{Under \texttt{-o pipefail}, pipelines' exit status will be the rightmost non-zero exit status if any command in the pipeline fails, as opposed to just yielding the exit status of the last command.}

On the other hand, a wide variety of alternative shells exist. Some of these are quite popular, like zsh (which became the default interactive shell on macOS starting in Catalina~\cite{macoszsh}) or fish; others are newer, slicker, more radical reimaginings of the shell (Elvish, nushell).
Some participants suggested that other languages could provide useful wisdom about techniques for evolution and replacement: C improvements, static analysis, and mooted “C killers” such as Go and Rust. Many such attempts at replacement end up simply “adding alongside”; it isn't clear whether any abstraction as pervasive as the Unix shell has ever truly been replaced in the history of computing.
Alternative shells, or a world where there's good support for more than one shell, bring about a ``two shells'' problem, which has occurred before with csh. As Chet Ramey pointed out, ksh was the effort to reunite the shell's programming and interactive features. People eventually get tired of changing back and forth unless there are  significant improvements one way or the other.

In discussions of usability (Section~\ref{sec:usability}), the shell's arcane quoting rules were a common concern. Should everything unquoted be a constant? Should field splitting be off by default? Should globbing be less dynamic? Any change here is guaranteed to cause issues with legacy scripts. Other languages limit these problems by constraining the form of various lexemes, especially identifiers. The shell's task is harder because what would be identifiers in many languages are often filenames in the shell, and these may use almost any character (except NUL and ‘/'). Peculiarities of the shell's use cases may also play a part: globbing is often used to recover structure post-hoc from a collection of names, whereas in another language the named things may have been created as a list to begin with. The lack of pointers per se, and reliance instead on computed (file)names, may also play a part.

Portability issues with the shell and its specification extend to the shell's ecosystem---the commands available and the filesystem.
The ambient commands are the shell's `library functions'.  Stephen Kell noted that arguably the shell doesn't need features for libraries or modularity, because its available library code is simply the commands on the current \texttt{PATH}, and more generally that extensibility can come from being situated within a filesystem.
Timothy Roscoe mentioned that the shell's success is tied to the success of Unix and its IPC and file systems---and therefore it is difficult to know what part of its success is inherent, or simply because we mostly use Unix now. Roscoe suggested that compositionality of processes might have led the shell's design instead of the other way around, and Chet Ramey agreed: ``the pipeline idea was quickly followed by new shell syntax, which was followed by an orgy of script editing to use it''.
Attempts at other shells that didn't `stick' may be accountable to failures of their ecosystem rather than of ideas for the shell itself. One such example is Rexx, which was developed for IBM 370 and had several interesting features like arbitrary fork/joins. Furthermore, any attempts at redesigning/improving the shell also have to look at Powershell, e.g., its native support for distribution.

\subsection{Outlook}

The two views of shell portability---bureaucratic (the POSIX spec) and pragmatic (various implementations)---are in conflict.
Clearly delineating goals and use cases for portability will help clarify the situation.
Thoughtful work on reconciling shell implementations must also consider reconciling shell ecosystems.

\section{Who attended?}

We list attendees in alphabetical order by last name. Some attendees on Zoom did not have a name set, and are omitted with our apologies.

Reto Achermann, Abutalib Aghayev, Sid Agrawal, Marcos Aguilera, Carol Alexandru, Ardalan Amiri Sani, Vaastav Anand, Vinay Banakar, Subho Banerjee, Andrew Baumann, Nishant Bhaskar, Nathan Bronson, Joao Carreira, Andy Chu, Jurgen Cito, David Cock, Tom Van Cutsem, Jackson Dagger, Ting (Dustin) Dai, Christos Dimoulas, Marco Faltelli, Georgia Fragkouli, Ian Friedman, Michał Gajda, Elena Glassman, Anitha Gollamudi, Sathish Gopalakrishnan, Sam Grayson, Marco Guazzone, Michael Greenberg, Arjun Guha, Aditya Gupta, A. Finn Hackett, Benjamin Hindman, Christopher Hodsdon, Vidar Holen, David Holland, Shayan Hosseini, Nora Hossle, Yao Hsiao, Lukas Humbel, Liz Izhikevich, Konstantinos Kallas, David R Karger, Baris Kasikci, Kim Keeton, Stephen Kell, Nodir Kodirov, Eddie Kohler, Erik Krogen, Vaibhav Kurhe, Fadhil I. Kurnia, Andrea Lattuada, Julia Lawall, Hongyi Liu, Tony Mason, Doug McIlroy, Hongyu Miao, Samantha Miller, Jeff Mogul, Derek Murray, Brad N. Karp, Akshay Narayan, Joel Nider, Øyvind Nohr, Sruthi P.C, Nathan Pemberton, Solal Pirelli, Ruzica Piskac, Stratos Psomadakis, Deepti Raghavan, Alagappan Ramanathan, Chet Ramey, Geoff Ramseyer, Ashwin Rao, Timothy (Mothy) Roscoe, Larry Rudolph, Ahmed Saeed, Alireza Sanaee, Johann Schleier-Smith, Malte Schwarzkopf, Ishan Sharma, Bingyu Shen, Jiasi Shen, Robert Soule, Diomidis Spinellis, Caleb Stanford, Brent Stephens, Lilia Tang, Geoffrey Thomas, Sam Tobin-Hochstadt, Amin Tootoonchian, Muhammed E. Ugur, Antero Vainio, Nikos Vasilakis, Ymir Vigfusson, John Wilkes, Keith Winstein, Le Xu, Tianyin Xu, Chen-yu (Robin) Yen, Matei Zaharia, Niloofar Zarif, Irene Zhang, Tianyi Zhang, Wen Zhang, Danyang Zhuo, and Gefei Zuo.

\section{Acknowledgments}

We are grateful, of course, to the panel's participants---and especially to Eddie Kohler and Baris Kasikci for organizing HotOS.
We owe a special thanks to our panelists, Arjun Guha, Deepti Raghavan, Chet Ramey, and Diomidis Spinellis.
Chet Ramey provided many insightful comments on this report;
Ting Dai and Solal Pirelli provided helpful comments on an early draft;
and Doug McIlroy offered fixes and insightful comments.

\bibliography{shell}

\end{document}